\documentclass[aps,preprint,nofootinbib,preprintnumbers,superscriptaddress,eqsecnum]{revtex4-1}

\usepackage{color}

\newcommand{\nb}[1]{\color{blue}}

\newcommand{\hl}[1]{\color{magenta}}

\usepackage[
	  pagebackref=false,
	  colorlinks=true,
      linkcolor=blue,
      urlcolor=blue,
      filecolor=black,
      citecolor=red,
      pdfstartview=FitV,
      pdftitle={},
        pdfauthor={Paolo Glorioso, Michael Crossley, Hong Liu},
        pdfsubject={},
        pdfkeywords={},
        pdfpagemode=None,
        bookmarksopen=true
      ]{hyperref}

\usepackage[normalem]{ulem}
\usepackage{amsmath}
\usepackage{enumerate}
\usepackage{amsfonts}
\usepackage{epsfig}
\usepackage{mathbbol}

\def\Im{\mathop{\rm Im} }
\def\Re{\mathop{\rm Re} }

\newcommand\p{\ensuremath{\partial}}

\newcommand\vev[1]{{\ensuremath{\left\langle{#1}\right\rangle}}}

\newcommand{\be}{\begin{equation}}
\newcommand{\ee}{\end{equation}}
\newcommand{\bea}{\begin{eqnarray}}
\newcommand{\eea}{\end{eqnarray}}
\newcommand{\bega}{\begin{gather}}
\newcommand{\eega}{\end{gather}}

\newcommand{\bi}{\begin{itemize}}
\newcommand{\ei}{\end{itemize}}
\newcommand{\ben}{\begin{enumerate}}
\newcommand{\een}{\end{enumerate}}
\newcommand{\bca}{\begin{cases}}
\newcommand{\eca}{\end{cases}}
\newcommand{\bln}{\begin{align}}
\newcommand{\eln}{\end{align}}
\newcommand{\bst}{\begin{split}}
\newcommand{\est}{\end{split}}
\def\ie{\begin{equation}\begin{aligned}}
\def\fe{\end{aligned}\end{equation}}
\newcommand{\bma}{\le(\begin{matrix}}
\newcommand{\ema}{\end{matrix}\ri)}

\newcommand\al{{\alpha}}
\def\b{{\beta}}
\newcommand\ep{\epsilon}
\newcommand\sig{\sigma}

\newcommand\lam{\lambda}

\newcommand\De{{\ensuremath{{\Delta}}}}

\newcommand\ka{\kappa}

\newcommand\Lra{{\Longrightarrow}}
\newcommand\ov{\over}

\newcommand\apr{{\ensuremath{{\alpha'}}}}
\def\le{\left}
\def\ri{\right}

\newcommand\sN{{\ensuremath{{\mathcal N}}}}

\begin{document}

\title{On systems of maximal quantum chaos}

\author{Mike Blake}
\affiliation{School of Mathematics, 
University of Bristol, Fry Building, \\Woodland Road, Bristol BS8 1UG, UK }

\author{Hong Liu}
\affiliation{Center for Theoretical Physics, 
Massachusetts
Institute of Technology, \\
Cambridge, MA 02139 }

\preprint{MIT-CTP/5283}

\begin{abstract}
A remarkable feature of chaos in many-body quantum systems is the existence of a bound on the quantum Lyapunov exponent. An important question is to understand what is special about maximally chaotic systems which saturate this bound. Here we provide further evidence for the `hydrodynamic' origin of chaos in such systems, and discuss 
hallmarks of maximally chaotic systems. We first provide evidence that a hydrodynamic effective field theory of chaos we previously proposed should be understood as a theory of maximally chaotic systems. We then emphasize and make explicit a signature of maximal chaos which was only implicit in prior literature, namely the suppression of exponential growth in commutator squares of generic few-body operators.  We provide a general argument for this suppression within our chaos effective field theory, and illustrate it using SYK models and holographic systems. We speculate that this suppression indicates that the nature of operator scrambling in maximally chaotic systems is fundamentally different to scrambling in non-maximally chaotic systems. We also discuss a simplest scenario for the existence of a maximally chaotic regime at sufficiently large distances even for non-maximally chaotic systems.
\end{abstract}

\today

\maketitle

\tableofcontents

\section{Introduction}

One of the key dynamical concepts characterizing  a quantum many-body system is  
scrambling of quantum information. Suppose we slightly disturb a system by inserting a local few-body operator.
Under time evolution, the operator will grow in physical and internal spaces (if there is a large number of single-site degrees of freedom), during which quantum information of the original disturbance is scrambled among more and more degrees of freedom. While for a general many-body system such evolutions are extremely complicated,
it has been increasingly recognized during the last decade that there are remarkable universalities in this process \cite{Aleiner,Roberts:2014ifa,Swingle,Nahum:2017yvy,Khemani:2017nda,Chowdhury:2017jzb, Shenker1, Roberts, Shenker2, MSS, MS,Roberts:2018mnp, Jensen, MSY, Kitaev}. For example, out-of-time-ordered correlators (OTOCs) have emerged as important 
probes of scrambling in chaotic systems. In particular, in chaotic systems with a large number of 
single-site degrees of freedom, the commutator squares of 
generic few-body operators grow exponentially with time at early times (below $\sN$ characterizes the number of 
single-site degrees of freedom)
\be \label{heni}
\vev{[W(t), V(0)]^2} \propto {1 \ov \sN} e^{\lam t}
\ee
with a quantum Lyapunov exponent $\lam$ bounded\footnote{For notational consistency with \cite{Blake:2017ris}, we denote the inverse temperature by $\beta_0$. Henceforth we set $\hbar = 1$.} by~\cite{MSS}
\be \label{maxb}
\lam \leq \lam_{\rm max} = {2 \pi \ov \hbar \b_0}  \ .
\ee
The bound, which has no classical analogue, follows from unitarity and analyticity. 

A variety of systems which saturate the bound have also been found, including holographic systems, SYK and its variations,
two-dimensional CFTs with a large central charge (and a gap), CFT correlation functions in the light-cone limit, and so on \cite{Shenker1,Roberts, MSY, MS, Jensen, Kitaev, Gu, Davison, Turiaci:2016cvo, Haehl:2018izb, Haehl:2019eae}. 
What are special about these maximally chaotic systems? One common theme that has emerged in such systems is that finite temperature dynamics of stress tensor appears to underlie the observed maximally chaotic behavior. 
In other words, the chaotic behavior appears to be ``hydrodynamic'' in nature. 

It is tempting to use the hydrodynamic nature of chaos as a defining feature of maximally chaotic systems. In this paper we would like to present further support for this idea, and to discuss a few general properties of maximally chaotic systems. 

 In~\cite{Blake:2017ris}, we proposed an effective field theory for describing chaotic behavior based on ``quantum'' hydrodynamics.  While the theory was motived by maximally chaotic systems, at the time it was not completely clear whether it works only for maximally chaotic systems. In this paper we first present a simple argument which shows that a system described by such an EFT must be maximally chaotic. We thus find strong support, if not a proof, that for chaos to be hydrodynamic in nature requires the system to be maximally chaotic. It also follows that various implications of the EFT discussed in~\cite{Blake:2017ris} 
should be considered as distinctive features of maximally chaotic systems. {For instance pole-skipping in the energy-density two-point function, first discussed in \cite{Saso,Blake:2017ris}, has been argued to be a `smoking-gun' signature of the hydrodynamic nature of chaos and has since been shown to occur in a wide-variety of maximally chaotic systems \cite{Blake:2018leo,Grozdanov:2018kkt,Haehl:2018izb,Haehl:2019eae,Ahn:2019rnq,Ahn:2020bks,Ramirez:2020qer}. See also~\cite{Choi:2020tdj} for a nice discussion of how pole-skipping generalises to non-maximally chaotic systems.

In this paper we will highlight 
another implication of the EFT: the exponential growth in the commutator square~\eqref{heni} 
is suppressed at leading order for maximally chaotic systems. We also illustrate the suppression of the commutator square in SYK models and holographic theories\footnote{The suppression of exponential growth in commutator squares in these models is implicit in the results presented in \cite{Kitaev, kitaIAS, Gu:2018jsv, Shenker2}.}. In particular we give a general argument showing that in holographic theories the commutator square is determined at leading order in $1/{\cal N}$ by stringy corrections\footnote{The importance of stringy corrections in determining the form of the commutator square in SYK models is discussed in \cite{Gu:2018jsv}.}.

 It has been observed in various examples that even non-maximally chaotic systems could exhibit maximally chaotic behavior 
 at sufficiently large distances~\cite{Shenker2, Gu, Gu:2018jsv, Mezei:2019dfv}. For such systems, the chaos EFT can apply at large distances. We discuss a simplest scenario for such a phenomenon.

The plan of the paper is as follows. In Sec.~\ref{sec:eft} we give a quick review of key features of the chaos EFT. 
In Sec.~\ref{sec:max} we show that the EFT describes systems with maximal chaos. 
In Sec.~\ref{sec:supp} we discuss suppression of the commutator square~\eqref{heni}. 
In Sec.~\ref{sec:velo} we discuss a simplest scenario for the existence of a maximally chaotic regime at sufficiently large distances for non-maximally chaotic systems. Sec.~\ref{sec:conc} concludes with a brief discussion of future questions. Appendix~\ref{app:scattering} contains some technical details for Sec.~\ref{sec:supp}.

\section{A quick review of the EFT for maximal chaos}
\label{sec:eft}

In this section we present a quick review of the chaos EFT introduced in~\cite{Blake:2017ris}, which is 
based on the following elements (to avoid cluttering notations 
below we first use $0+1$ dimensional systems as an illustration): 

\ben 

\item  One imagines the scrambling of a generic few-body operator $V(t)$ at a finite temperature allows a coarse-grained description in terms of building up a ``hydrodynamic cloud,'' i.e. we can write 
\be \label{gco}
V(t) = V [\hat V, \sig (t)] \ .
\ee  
Here $\hat V (t)$ is a ``bare'' operator involving the original degrees of freedom of $V$, and $\sig (t)$ is a chaos mode which 
describes the growth of the operator in the space of degrees of freedom. Bare operators which involve different degrees of freedom have no correlations, i.e. $\vev{\hat V \hat W} =0$ for generic $V \neq W$.

\item One identifies the chaos mode $\sig$ with the collective degrees of freedom associated with energy conservation, and thus 
its dynamics is governed by hydrodynamics. However, in  order to capture the growth of $V(t)$, one needs a 
``quantum hydrodynamics'' which is valid in the regime 
$\De t \sim \b_0$, with $\De t$ typical time scales of interests. This quantum hydrodynamic theory can be considered as a generalization of the conventional hydrodynamics without doing a derivative expansion\footnote{Recall that conventionally hydrodynamics is formulated as a derivative expansion and is 
valid for $\De t \gg \b_0$.}, and is non-local. Its action, which also incorporates dissipative effects, can be written using the techniques developed in~\cite{CGL,CGL1}. The explicit form of the action $S_{\rm EFT} [\sig]$  is not important for our discussion below (we refer interested readers to~\cite{Blake:2017ris} for details) except that the Lagrangian depends on $\sig$ only through derivatives and 
the equilibrium solution is given by
\be \label{ejn}
\sig_{\rm eq} (t) = t  \ .
\ee

\item To describe a chaotic system, we still need to impose an additional shift symmetry 
\be \label{oem1}
e^{-\lam \sig} \to e^{-\lam \sig} + c
\ee
on the action 
$S_{\rm EFT} [\sig]$ and on the coupling~\eqref{gco} of $\sig$ to general few-body operators.
Here $c$ is an arbitrary {positive} constant and $\lam$ is the quantum Lyapunov exponent. Near equilibrium we introduce  
the deviation of $\sig (t)$ from the equilibrium solution~\eqref{ejn} 
\be\label{hen0}
\ep (t) = \sig (t) - t 
\ee
For infinitessimal $\epsilon(t)$ the symmetry \eqref{oem1} implies a symmetry under
\be 
\ep \to \ep - c e^{\lam t}  \ .
\ee
This shift symmetry results in an exponential growing piece in two-point functions of $\sigma$ which is the origin of chaotic behaviour within the EFT.  Nevertheless the EFT description of $\sigma$ appears to be self-consistent for any value of $\lam$. In particular, stress-tensor correlation functions do not contain an exponentially growing piece, and fluctuation-dissipation relations of the stress tensor  do not put any constraint on the value of~$\lam$. 

 We will, however, show in Sec.~\ref{sec:max} that when coupled to a generic operator through~\eqref{gco}, fluctuation-dissipation relations of the operator require $\lam$ to take the maximal value~\eqref{maxb}.

\begin{figure}
\begin{center}
\resizebox{60mm}{!}{\includegraphics{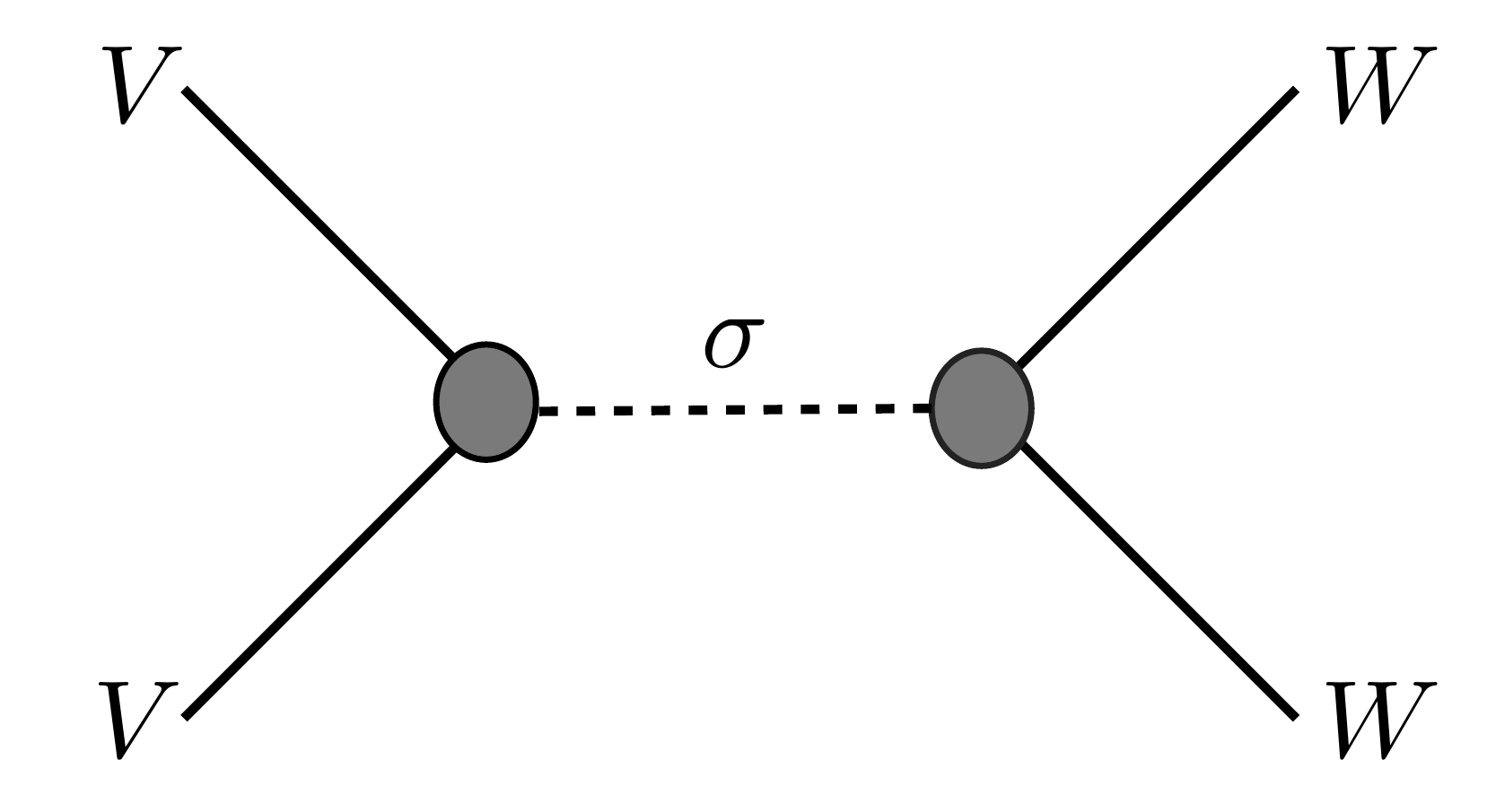}}
\caption{At leading order in large ${\cal N}$ correlation functions are controlled by exchange of hydrodynamic fields $\sigma(t)$. 
}
\label{fig:sigmainteraction}
\end{center}
\end{figure}

\item Using $S_{EFT}[\sig]$ and the coupling~\eqref{gco} one can calculate correlation functions of generic operators perturbatively in $1/{\cal N}$. To leading order in $1/{\cal N}$ it is sufficient to consider tree-level exchange of $\sigma$ as in Fig.~\ref{fig:sigmainteraction}. We require the coupling~\eqref{gco} between $\sig$ and a general operator be also invariant under the shift symmetry~\eqref{oem1}. Near equilibrium, equation~\eqref{gco} can be written using~\eqref{hen0} in a form 
\be \label{cop}
V (t) = \hat V (t) + L^{(1)} [\hat V \ep ] (t)  + O(\ep^2)   
\ee
where $L^{(1)}$ is some differential operator acting on both $\hat V$ and $\ep$.  Invariance under the shift symmetry implies the condition  
\be \label{om}
 L^{(1)}_{t_1} [g_V  (t_{12})  e^{\lam t_1}]   + L^{(1)}_{t_2} [g_V  (t_{12})  e^{\lam t_2}] = 0
 \ee
 where subscript $t$ in $L^{(1)}_t$ refers to the variable of the derivatives and $g_V(t)$ is the two-point function of $\hat{V}$
 \be
 \label{gvdef}
 g_V(t) = \langle \hat{V}(0) \hat{V}(t) \rangle  = G^{\hat{V}}_-(t)
 \ee
 It was also found in \cite{Blake:2017ris} that in order for the exponentially growing behavior to cancel for all possible configurations of TOC it is also necessary to impose the version of~\eqref{om} with $\lam \to -\lam$, i.e. 
\be \label{omm2}
 L^{(1)}_{t_1} [g_V  (t_{12})  e^{-\lam t_1}]   + L^{(1)}_{t_2} [g_V  (t_{12})  e^{-\lam t_2}] = 0 \ .
 \ee

\een
The shift symmetry may be considered as an emergent ``macroscopic'' symmetry.\footnote{Here by ``macroscopic'' we refer to scales of order $\b_0$.} It has many implications: (i)  constrains the structure of correlation functions to all derivative orders; (ii) implies existence of an exponentially growing mode the exchange of which leads to exponential growth of OTOCs;  (iii) leads to a connection between energy diffusion and the butterfly velocity~\cite{Patel, MB, MB2, MB3, MB4, Gu, Davison}. Furthermore, the dual role of $\sig$ as the mode for ``propagating chaos'' and energy conservation leads to a new phenomenon called ``pole-skipping"~\cite{Saso, Blake:2017ris} in the energy-density two-point function, which has been confirmed in a variety of maximally chaotic systems including general holographic systems dual to Einstein gravity and its higher derivative extensions~\cite{Blake:2018leo,Grozdanov:2018kkt,Haehl:2018izb,Haehl:2019eae}.

Let us now elaborate a bit more on~\eqref{cop}--\eqref{omm2}, and how to use 
the above elements to compute four-point functions of generic few-body operators. We can expand $L^{(1)}$ as
\be \label{uen}
L^{(1)} [\hat V \ep ] = \sum_{n,m=0}^\infty c_{nm} \p_t^n \hat V \p^m_t \ep + {\cal O}{(\epsilon^2)} 
\ee
where $c_{mn}$ are constants. Then equation~\eqref{om} can be written more explicitly as 
\be \label{mme}
{F_{\rm even} (\lambda, t) \ov F_{\rm odd} (\lambda, t)} = - \tanh {\lam t \ov 2} \ .
\ee
where we have defined 
\be 
F_{\rm even} (\lambda, t) = \sum_{n \; {\rm even}} f_n (\lam) \p^{n}_t g_V (t) ,\qquad 
F_{\rm odd} (\lambda, t) = \sum_{n \; {\rm odd}} f_n (\lam) \p^{n}_t g_V (t)  \ .
\ee
with
\be 
f_n = \sum_m c_{nm} \lam^m  \ .
\ee
Similarly~\eqref{omm2} implies 
\be \label{mme1}
{F_{\rm even} (-\lambda, t) \ov F_{\rm odd} (-\lambda, t)} =  \tanh {\lam t \ov 2} \ .
\ee

\begin{figure}
\begin{center}
\resizebox{80mm}{!}{\includegraphics{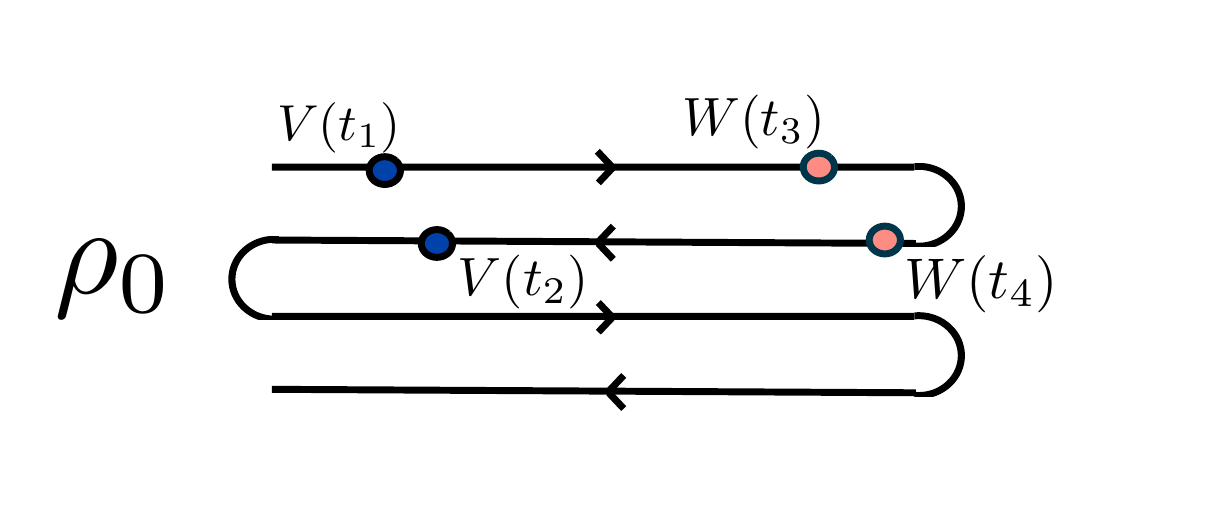}}
\resizebox{80mm}{!}{\includegraphics{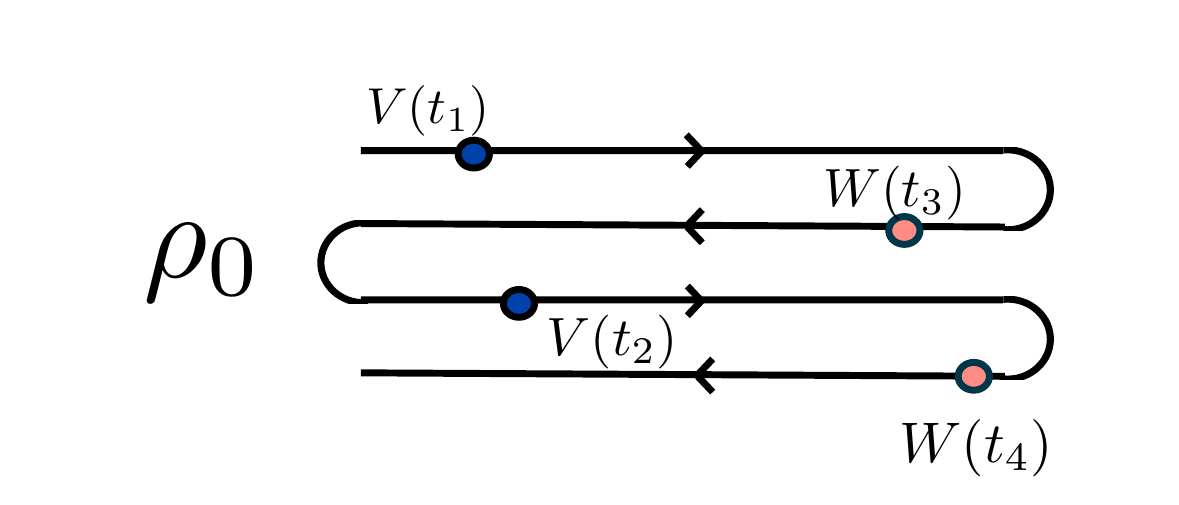}}
\caption{Left: operator insertions for~\eqref{o1}. Right: operator inserations for~\eqref{o2}.
}
\label{fig:4contours}
\end{center}
\end{figure}

Four-point functions of generic few-body operators can be computed at leading order in $1/\sN$ expansion by tree-level exchange of $\sig$ as indicated in Fig.~\ref{fig:sigmainteraction}. 
For illustration, consider  two explicit examples indicated in Fig.~\ref{fig:4contours}. The first is a TOC\footnote{Throughout this paper we use the notation $\vev{O} = \mathrm{Tr(O \rho_0)}$ with $\rho_0 = e^{- \beta_0 H}/Z$ the thermal density matrix and take $V,W$ to be Hermitian.}
\be
\label{o1}
G_4(t_1,t_2,t_3,t_4) 
= \vev{V(t_2) W(t_4) W(t_3) V(t_1)  }
\ee
and the second is an OTOC 
\be \label{o2}
H_4(t_1,t_2,t_3,t_4) 
= \vev{W (t_4) V (t_2) W (t_3) V (t_1)}  
\ee
with $t_{3,4} \gg t_{1,2}$.  Using~\eqref{cop} we find~\eqref{o1} can be written as\footnote{Note that our definitions of $G_4, H_4$ differ slightly from those in \cite{Blake:2017ris} as we are not normalising them by the bare correlation function.}
\bega 
G_4 - g_V(t_{12}) g_W(t_{34})=  \le( L^{(1)}_{t_1}  L^{(1)}_{t_3}  G_+ (t_{31})  + L^{(1)}_{t_1}  L^{(1)}_{t_4} G_+ (t_{41}) + \ri.  \cr
\le. L^{(1)}_{t_2}  L^{(1)}_{t_3} G_- (t_{32}) +  L^{(1)}_{t_2}  L^{(1)}_{t_4} G_- (t_{42})  \ri) g_V  (t_{12}) g_W  (t_{34})
 \label{timeO}
\end{gather}
where $g_V ,g_W$ are two point functions of the bare operators $\hat V, \hat W$, $t_{12} = t_1 - t_2$, and 
in the above equation it should be understood that $L^{(1)}$'s also act on $g_V$ or $g_W $ after them. 
$G_\pm$ are Wightman functions\footnote{For $\lambda = \lambda_{\mathrm{max}}$ there are also additional exponential terms in $G_{\pm}$ of the form $t e^{\lambda t}$.} of $\ep(t)$, 
\bega
G_+(t) = \vev{\ep(t) \ep(0)} =  c_+ e^{\lambda t} + \cdots \;\;\;\;\;\;\; G_{-}(t) = \vev{\ep(0) \ep(t)} = c_- e^{\lambda t} + \cdots , \\ 
\label{delta}
\Delta(t) = \vev{ [\epsilon(t), \epsilon(0) ] } = -i c e^{\lambda t} + \cdots, \quad c = i (c_+ - c_-) 
\end{gather}
with $c_\pm$ some constants proportional to $1/{\cal N}$, and $c$ a real constant. Note that in the above expressions we have only made manifest the exponential terms in $t$. 
A similar expression can be obtained for $H_4$. Note that 
\be
\label{eq:difference}
H_4 - G_4 = \vev{[W(t_4),V(t_2)] W(t_3) V(t_1)} = L^{(1)}_{t_2}  L^{(1)}_{t_4} [g_V(t_{12}) g_W (t_{34})\De (t_{42}) ]\
\ee

One can check that due to~\eqref{om}--\eqref{omm2}, the exponentially growing terms are canceled in TOC~\eqref{timeO}. Then due to~\eqref{eq:difference} and~\eqref{delta} one finds that OTOC $H_4$ must contain an exponentially growing piece.

\section{Consistency of EFT requires maximal chaos} \label{sec:max}

In this section we present a simple argument to show that the consistency of the shift symmetry of the 
effective vertex and fluctuation-dissipation relations of general operators require that the Lyapunov exponent be maximal. 

In particular the two-point function $g_V$ of $V$ should satisfy the fluctuation-dissipation theorem (FDT), which can be used to constrain $\lam$ through~\eqref{mme}.  Applying FDT to $g_V$ in ~\eqref{gvdef} implies that $g_V(t) = g_V(i \beta_0 - t)$ (for ${\rm Im} \, t \in[0, \b_0)$) which in turn leads to 
\be \label{fdt1}
F_{\rm even} (\lam, t) = F_{\rm even} (\lam, i \b_0 - t), \qquad F_{\rm odd} (\lam, t) = - F_{\rm odd} (\lam,  i \b_0 - t) \ .
\ee
Assuming that equation~\eqref{mme} can be analytically continued to the range ${\rm Im} \, t \in [0, \b_0)$, we find that it is compatible with~\eqref{fdt1} only if 
\be 
\tanh {\lam t \ov 2} = \tanh {\lam (t - i \b_0) \ov 2}  \quad \Lra \quad \lam = {2 \pi \ov \b_0} \ 
\ee
We therefore conclude that these conditions are only consistent for a maximal Lyapunov exponent, and that the hydrodynamic theory proposed in \cite{Blake:2017ris} is a theory of maximally chaotic systems. 

A further connection to recent discussions of maximal chaos can be seen by noting that one can use the shift symmetry \eqref{mme} and \eqref{mme1} to write the exponentially growing behaviour in $H_4$ in a particularly simple form
\begin{equation}
\label{eq:OTOCexp}
H_4 -g_V(t_{12}) g_W(t_{34}) = i c \frac{F_{\mathrm{even}}(-\lambda, t_{12} )} {\mathrm{sinh}{\lambda t_{12} \ov 2}}\frac{{\tilde F}_{\mathrm{even}}(\lambda, t_{34} )}{\mathrm{sinh}{\lambda t_{34} \ov 2}} e^{\lambda {(t_3 + t_4- t_1 - t_2 )}/2} + \cdots \ .
\end{equation}
Where $\tilde{F}_{\mathrm{even}}(\lambda, t)$ is analogous to $F_{\mathrm{even}}(\lambda, t)$ defined above, but is defined using $g_W(t)$. The expression \eqref{eq:OTOCexp}  matches an ansatz for the OTOC of 0+1 dimensional systems proposed in \cite{Kitaev, kitaIAS}} in the case $\lambda = \lambda_{\mathrm{max}}$.

\section{Suppression of commutator square for maximally chaotic systems} \label{sec:supp} 

In this section we first present a general argument showing that in the chaos EFT, the commutator square~\eqref{heni} always vanishes at leading order. We then support the conclusion with another argument, and illustrate it 
using the examples of SYK model and holographic systems. In particular we note that in holographic systems the commutator square is always determined by stringy corrections. 

\subsection{EFT derivations} 

Now that we have shown that the shift symmetry requires maximal chaos, here we present a new prediction of this theory for the commutator square 
\begin{equation}
\label{commutatorsquare}
C_4(t_1,t_2,t_3,t_4) = \langle [W(t_4), V(t_2)] [W(t_3) ,V(t_1)] \rangle  \ .
\end{equation}
We are interested in the regime  $t_3, t_4 \gg t_1, t_2$.  The double commutator consists of two time ordered and two out-of-time ordered correlation functions, 
\begin{equation}
C_4(t_1,t_2,t_3,t_4) = H_4 + \tilde{H}_4 - G_{4} - \tilde{G}_{4}, 
\end{equation}
with $G_4, \tilde{G}_4$ time ordered
\begin{equation}
G_4 = \vev{V(t_2) W(t_4) W(t_3) V(t_1)}  \;\;\;\;\;  \tilde{G}_4 = \vev{W(t_4) V(t_2) V(t_1) W(t_3) }  \ ,
\end{equation}
and $H_4, \tilde{H}_4$ out-of-time ordered
\begin{equation}
\label{OTOCdef}
H_4 = \vev{W(t_4) V(t_2) W(t_3) V(t_1)}  \;\;\;\;\;  \tilde{H}_4 = \vev{V(t_2) W(t_4) V(t_1) W(t_3) }   \ .
\end{equation}
Using equations~\eqref{eq:OTOCexp},~\eqref{fdt1}, $\tilde H_4 (t_1, t_2, t_3, t_4)= H_4 (t_2 + i \b_0, t_1, t_3, t_4)$ and that $\lambda = \lambda_{\mathrm{max}}$ one finds
that the exponential pieces in $H_4$ and $\tilde H_4$ precisely cancel, i.e. $C_4(t_1, t_2, t_3, t_4)$ does not have any 
exponentially growing pieces. In fact, the contribution from tree-level exchange of $\sig$ 
to the double commutator in \eqref{commutatorsquare} is identically zero. 
Recall from~\eqref{eq:difference} 
\be 
H_4 - G_4 =  L^{(1)}_{t_2}  L^{(1)}_{t_4} [g_V(t_{12}) g_W (t_{34}) \De (t_{42}) ], \qquad \De (t_{42} ) = \vev{[\ep (t_4) , \ep (t_2)]} \ .
\ee
Likewise it is straightforward to show that
\bega 
\tilde{H}_4 - \tilde{G}_4 = - L^{(1)}_{t_2}  L^{(1)}_{t_4} [g_V(t_{12}) g_W (t_{34}) \De (t_{42}) ] = 
- (H_4 - G_4), \\  
\Lra \quad  C_4(t_1,t_2,t_3,t_4) = 0, \quad \text{at leading order in} \, {1 \ov \sN} \ .
\label{exactcancellation}
\end{gather}

\begin{figure}
\begin{center}
\resizebox{140mm}{!}{\includegraphics{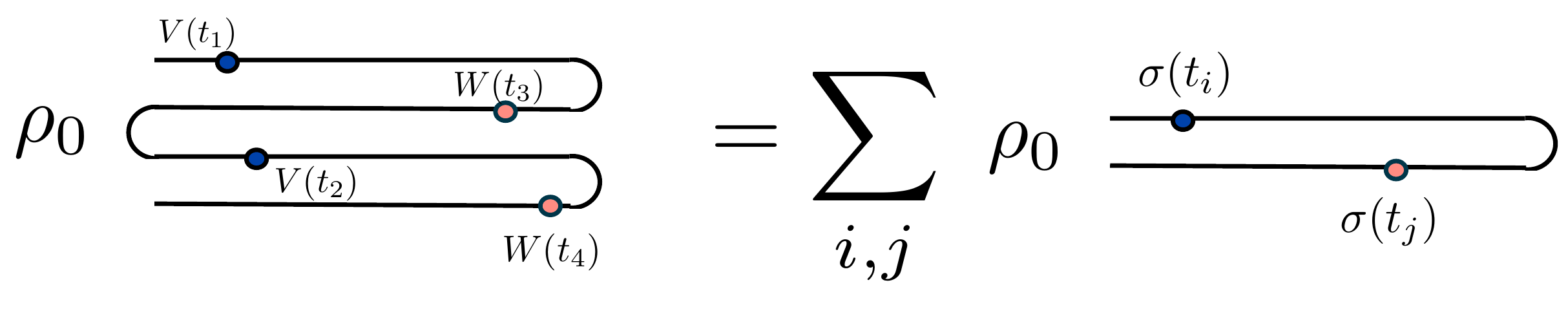}}
\caption{At leading order in large $\sN$ OTO four-point functions defined on a four-contour reduce to a sum of two-point functions of $\sigma(t)$. They can therefore be calculated from the effective action of the hydrodynamic field $\sigma$ on a CTP contour.}
\label{fig:otoreduction}
\end{center}
\end{figure}
The vanishing of~\eqref{commutatorsquare} at leading order can be intuitively understood from  Fig.~\ref{fig:otoreduction}: 
four-point functions all reduce to two-point functions of $\sig$, while for~\eqref{commutatorsquare} to be non-vanishing 
one needs four-point function of $\sig$, which are higher orders in $1/\sN$. 

We can also consider commutators which  are separated  in imaginary time, such as 
\begin{equation}
\label{commutatorsquarecomplex}
C_{4}^{\theta}(t_1,t_2,t_3,t_4) = \langle [W(t_4 + i \theta), V(t_2 + i \theta)] [W(t_3) ,V(t_1)] \rangle
\end{equation}
where in order for the double commutator to be defined it is necessary to use a configuration where imaginary parts of $t_2,t_4$ match and likewise for $t_1, t_3$.  The same arguments as above show that $ C_4^{\theta}(t_1,t_2,t_3,t_4) = 0$ at leading order in $1/{\cal N}$. 

While we have focused on $(0+1)$-dimensional systems for simplicity, we expect that in higher dimensional systems, 
the argument for the exact cancellation of the chaotic mode contribution to the commutator square $\langle [W(t_4,\vec{x}_4), V(t_2,\vec{x}_2)] [W(t_3,\vec{x}_3) ,V(t_1,\vec{x}_1)] \rangle $ also applies. 
Explicit examples of higher-dimensional chaos EFT include those for SYK chains \cite{Gu} and CFT constructions \cite{Haehl:2019eae, Haehl:2018izb}.

 Known examples of maximal chaos all happen in some limit, in addition to $\sN \to \infty$.
For the SYK model this corresponds to the low temperature limit, for holographic systems it is the classical gravity limit (or infinite coupling), and for CFTs in a Rindler wedge it is the light-cone limit. In such a limit, the stress tensor exchange gives the dominant contribution to OTOCs, and is responsible for the appearance of maximal chaos. 
For convenience of discussion we will denote the maximal chaos limit in these theories collectively as the $g \to \infty$ limit, with $g$ standing for the corresponding parameter in each system. Given that the contribution of the stress tensor exchange to the commutator square~\eqref{commutatorsquare} vanishes, the leading order result is then given by contributions from other modes. 
 We will discuss in Sec.~\ref{sec:exam} the examples of SYK and holographic systems in detail. In the conclusion Sec.~\ref{sec:conc} we also briefly speculate on the physical interpretation of the vanishing of the commutator square at leading order.
Here we will briefly comment on the general features. 

Including the contributions from an infinite number of other modes leads to two main corrections to the behavior of the OTOC. Firstly, the Lyapunov exponent is shifted 
\be \label{heni1}
\lam ={2 \pi \ov \b_0} - {c_1 \ov g^{\al_1}}, \quad g \to \infty  
\ee
where $c_1> 0$ and $\al_1> 0$ are some constants which depend on specific systems. Secondly, the prefactor to $e^{\lam t}$ in the OTOC also receives high order corrections in $1/g$. For concreteness consider the configuration $t_1 = 0$, $t_2 = i \epsilon$, $t_3 = t$, $t_4 = t + i \epsilon$ with $\epsilon \ll \beta_0$.  Then in the examples of SYK and holographic systems \cite{Shenker2, Kitaev, kitaIAS, Gu:2018jsv}, the OTOC has the form
\be 
\label{thermal}
H_4  (t) = i C (g) e^{- \frac{i c_1 \beta_0}{4 g^{\al_1}}}  e^{\lam t} = C(g) e^{i \lambda \beta_0/4}e^{\lam t} 
\ee
where 
$C(g)$ is real.
Plugging in the above expression into the commutator square\footnote{Note for this configuration $C_4^{\epsilon}(t) = 2 \Re (H_4(t)) + \dots$ with the dots indicating non-exponential pieces.} we then find that to leading order in the large $g$ limit,
\be \label{leadS}
C_4^{\epsilon} (t) =  \frac{ C c_1 \beta_0}{2 g^{\al_1}}  e^{{2 \pi \ov \b_0} t}, \qquad g \to \infty \ .
\ee
The leading order behavior of the prefactor in the above equation depends on how $C (g)$ behaves as $g \to \infty$. For example, if $C$ is finite as $g \to \infty$ then the prefactor vanishes as $g^{-\al_1}$ in the large $g$ limit.

\subsection{Other arguments} 

Before discussing explicit examples, for completeness we mention some other arguments which suggest that 
the commutator square vanishes for maximal chaos at leading order. Consider first the commutator square $C_4^{-\beta_0/2}(t)$ corresponding to \eqref{commutatorsquarecomplex} with $t_1 = t_2 = 0$ and $t_3 = t_4 = t$ and $\theta = -\beta_0/2$. This can be written as \cite{MSS}
\begin{equation}
C_4^{-\beta_0/2}(t) = F(t + i \beta_0/4) + F (t - i \beta_0/4) + \cdots 
\end{equation}
with $F(t)$ the thermally regulated OTOC where operators are symmetrically placed around the unit circle 
\begin{equation}
\label{MSSotoc}
F(t) = \mathrm{Tr} \bigg(W(t) \rho_0^{1/4} V(0) \rho_0^{1/4} W(t) \rho_0^{1/4} V(0) \rho_0^{1/4} \bigg) \ .
\end{equation}
Assuming $F (t) \sim C e^{\lambda t}  $ with real\footnote{Note $F(t)$ is real.} $C \sim 1/{\cal N}$ we then have 
\begin{equation}
\label{supp}
C_4^{-\beta_0/2}(t) = 2 C \, \mathrm{cos}(\lambda \beta_0/4) e^{\lambda t} + \cdots \ .
\end{equation}
For maximal chaos we find a suppression as the prefactor factor $\mathrm{cos}(\lambda \beta_0/4)$ vanishes. 

More generally, in~\cite{Kitaev, kitaIAS}  an ansatz was proposed for the leading order in $1/{\cal N}$ exponential growth of the OTOC correlation function 
\begin{equation}
\label{ansatz}
H_4(t_1,t_2,t_3,t_4) = C e^{i \lambda \beta_0/4} e^{\lambda (t_3 + t_4 - t_1 - t_2)/2} \Upsilon^{R}_{\mathrm{WW}}(t_{34}) \Upsilon^{A}_{\mathrm{VV}}(t_{12})
\end{equation}
where the overall prefactor $C \sim 1/{\cal N}$ is real, and $\Upsilon^{R}_{\mathrm{WW}}(t_{34})$, $\Upsilon^{A}_{\mathrm{VV}}(t_{12})$ are advanced and retarded vertex functions that satisfy fluctuation-dissipation relations. The phase factor 
$e^{i \lambda \beta_0/4}$ is consistent with~\eqref{MSSotoc} being real. For $\lambda = 2\pi/\beta_0$, equation~\eqref{ansatz} is also consistent with 
the form~\eqref{eq:OTOCexp} from the effective theory.  
It then follows from~\eqref{ansatz} that 
\begin{equation}
\label{sepcomm}
C_4^{\theta}(t_1,t_2,t_3,t_4) = 2 C \; \mathrm{cos} (\lambda \beta_0/4) e^{ \lambda (t_3 + t_4 - t_1 - t_2)/2} \Upsilon^{\mathrm{A}}(t_{12} - i \theta)\Upsilon^{\mathrm{R}}(t_{34} - i \theta)  \ .
\end{equation}
Again for maximal chaos the $\mathrm{cos} (\lambda \beta_0/4)$ factor suppresses the leading contribution to $C_4^\theta$.

\subsection{Examples} 
\label{sec:exam}

We now use the examples of SYK model and holographic systems to illustrate the somewhat abstract discussion above. 
A suppression of exponential growth in the commutator square of various maximally chaotic systems has previously been observed in \cite{Kitaev, kitaIAS, Shenker2, Gu:2018jsv}.

\subsubsection{SYK model}

The ansatz~\eqref{ansatz} {for a general system} was in fact motivated in large part by the form of the OTOC in soluble (i.e. strong coupling or large-$q$) limits of the SYK model. For the SYK model, the parameter $g$ in~\eqref{heni} can be identified as $g = \b_0 J$ and $g \to \infty$ corresponds to the low temperature limit. In this case \cite{MS, Gu:2018jsv}
\be 
\lam = {2 \pi \ov \b_0} \le(1 - {c \ov \b_0 J} \ri)
\ee
i.e. $\al_1 =1$ in~\eqref{heni1}  and 
$C \sim {\b_0 J \ov \sN}$.  We thus find 
\be 
C \cos  (\lambda \beta_0/4)  \sim O((\b_0 J)^0) {1 \ov \sN} 
\ee
which is suppressed by a factor $1/(\b_0 J)$ compared with OTOC. These features can be illustrated explicitly in the large-$q$ limit, for which a closed form expression of $H_4$ for all $\b_0 J$ was obtained in \cite{Gu:2018jsv}. The exponential growing part of $H_4$ can be written as  
\be 
\label{largeq}
H_4 
\approx  \frac{1}{{\cal N} \cos (\nu \pi/2)} {  e^{{\pi \nu \ov \beta_0} (t_3 + t_4 - t_1 - t_2 + i \beta_0/2)} \ov  (2 \mathrm{cosh}(\frac{\pi \nu}{\beta_0}(t_{12} - \frac{i \beta_0}{2} )))(2 \mathrm{cosh}(\frac{\pi \nu}{\beta_0}(t_{34}- \frac{i \beta_0}{2}) ))} 
\ee
where $v \in (0,1)$ parametrises the strength of the coupling $\b_0 J$ through
\begin{equation}
\frac{\pi \nu}{ \cos (\pi \nu /2)} =\b_0 J \ .
\end{equation}
In the large $\b_0 J$ limit, $\nu \approx 1$ and the prefactor of $H_4$ is proportional to $ {\b_0 J \ov \sN}$. In contrast, the commutator OTOC is given by
\be 
\label{largeq}
C_4 
\approx  \frac{2}{{\cal N}} {  e^{{\pi \nu \ov \beta_0} (t_3 + t_4 - t_1 - t_2)} \ov  (2 \mathrm{cosh}(\frac{\pi \nu}{\beta_0}(t_{12} - \frac{i \beta_0}{2} )))(2 \mathrm{cosh}(\frac{\pi \nu}{\beta_0}(t_{34}- \frac{i \beta_0}{2}) ))} 
\ee
As indicated above the prefactor of the exponential growth is suppressed relative to $H_4$ by a factor $2 \cos ( \pi \nu /2) = 2 \cos(\lambda \beta_0/4) $, and is finite in the strong coupling limit $\beta_0 J \to \infty$.

Note that the leading order result for $H_4$ in the large $\b_0 J$ limit can also be obtained from the Schwarzian theory, which 
provides a specific example of the hydrodynamic effective theory discussed in Section~\ref{sec:eft}. 
From the argument in \eqref{exactcancellation} we expect that the non-exponential parts of the commutator square should also cancel identically in the Schwarzian theory. This can indeed be confirmed  explicitly from analytically continuing the results in \cite{MSY}.

\subsubsection{Holographic systems} 
\label{sec:holography}

As another example, let us consider the behavior of the commutator square in higher dimensional holographic theories, both in classical gravity and in string theory. Earlier discussion includes~\cite{Shenker2,Mezei:2019dfv}.  
For concreteness we consider the out-of-time-ordered correlator \begin{equation}
\label{otocholography}
H_4(t, \vec{x}) = \langle W (t + i \epsilon,\vec{x}) V(i \epsilon,0) W (t,\vec{x}) V(0,0) \rangle
\end{equation}
where $\epsilon \ll \beta_0$, and the corresponding commutator square
\begin{equation}
\label{commutatorcholography}
C_4(t, \vec{x}) = \langle [W (t + i \epsilon,\vec{x}),  V(i \epsilon,0)] [W (t,\vec{x}) , V(0,0)] \rangle \ .
\end{equation}

At the level of the gravity approximation, which corresponds to $\sN \to \infty$ and $\lam_h \to \infty$ in the boundary theory\footnote{For example, when the boundary theory is  the $\sN=4$ Super-Yang-Mills theory, $\lam_h$ is the 't Hooft coupling.}, the exponential growth of $H_4$ has the form \cite{Roberts, Shenker2} %
\be 
H_4 (t,\vec{x}) = i {c \ov \sN} e^{2 \pi/\beta_0 (t - |\vec{x}|/v_B)}
\label{gravityotoc} 
\ee
where $c \sim O(1)$ is real. 
The expression \eqref{gravityotoc} comes from graviton exchanges which translates into the boundary theory from exchanges of the stress tensor. Such exchanges are fully captured by our EFT and we thus expect that in the classical gravity 
\be 
C_4 = 0, \quad \sN \to \infty, \quad {\lam_h \to \infty} 
\label{holozero}
\ee
including the non-exponential parts.

 In Appendix~\ref{app:scattering} we show that within the eikonal approximation the sum of $H_4$ and $\tilde{H}_4$ vanishes identically in any theory dual to classical gravity. As such, there is no exponential growth in $C_4$ to leading order in $1/{\cal N}$ in classical gravity. In particular, we find that this cancellation is a direct consequence of the fact the phase-shift for the bulk gravitational scattering process is real. The suppression of the commutator OTOC in gravitational systems can therefore be seen to be a consequence of 
bulk scattering being elastic. 
 
The above considerations hold for any system dual to a classical gravity theory, even if the bulk theory contains higher derivative terms reflecting certain finite coupling corrections. In order to find a non-vanishing expression for $C_4$ at leading order in $1/{\cal N}$, it is necessary to include not just the effects of gravitational scattering but also the exchange of stringy modes at finite string coupling $\alpha'$. This translates into the boundary theory as exchanges of an infinite number of other intermediate operators (including in particular high-spin operators) in addition to the stress tensor. 
 
The effect of such stringy corrections on the OTOC  %
has been discussed in detail in~\cite{Shenker2}. The discussion is a bit technical, however the qualitative features of these results are illuminating. The important result is that $H_4$ is controlled by the phase shift of a bulk scattering process. 
 In particular, the functional form of the OTOC shows a sharp crossover between a stringy regime for which $|\vec{x}|/t < v_*$ and a graviton dominated regime for $|\vec{x}|/t > v_*$ where for the black hole background studied in \cite{Shenker2, Mezei:2019dfv} we have
 \begin{equation}
 v_* = \frac{d^2 \alpha'}{4 v_B R^2}, \;\;\;\;\;\;\;\;\;\;\;\;\;  v_B = \sqrt{\frac{d}{2(d-1)}}
 \end{equation}
 with $R$ the AdS curvature radius and $d$ the number of spacetime dimensions in the boundary theory. 

To elaborate, for short distances $|\vec{x}| < t/v_*$, where $v_*$ is some specific velocity, $H_4$ can be computed from an integral representation by a saddle point approximation and stringy corrections are important. In this regime it takes the form 
\be 
H_4 (t,\vec{x}) =  {c \ov {\sN t^{(d-1)/2}}} f_1(|\vec{x}|/t) e^{\lam t} e^{- \frac{|\vec{x}|^2}{4 D t}} \ . 
\label{stringotoc} 
\ee 
where $f_1 (|\vec{x}|/t)$ is complex, the string corrected Lyapunov exponent $\lam$ is given by 
\be 
\lam = {2 \pi \ov \b_0} \le(1 - {d (d-1) \apr \ov 4 R^2} \ri) \ .
\label{stringlyapunov}
\ee
and the constant $D$ is given by
\begin{equation}
\label{chaosdiff}
D = \frac{d^2 \alpha' \beta_0}{16 \pi R^2}
\end{equation}

In the regime $|\vec{x}| > t/v_*$  even at finite $\apr$, the integral for $H_4$ is dominated by a graviton pole. As a result 
the contribution from graviton exchange dominates, and $H_4$ is given simply by the classical gravity result \eqref{gravityotoc}.

Here we point out that, for $C_4$~\eqref{commutatorcholography},  such a crossover does not exist, and its behavior  is  controlled by stringy exchanges for all $t$ and $\vec{x}$.  This can be intuitively expected from that the gravitational contribution to $C_4$ vanishes identically to leading order in $1/{\cal N}$. 
We give technical details in Appendix~\ref{app:scattering}.  There we show that the 
form of $C_4$ at large ${\cal N}$ is controlled by the imaginary part of the phase shift for a bulk scattering, which is zero for graviton exchange, but non-zero due to the effects of stringy modes. As a result, there is no graviton pole in the integral expression for $C_4$
and  there is no cross-over in its behavior. More explicitly, as shown in Appendix~\ref{app:scattering}, for all $|{\vec x}|/t$, $C_4$ has the form  
\begin{eqnarray} \label{oen}
C_4 (t,\vec{x}) &=& {c \ov {\sN t^{(d-1)/2}}}f_2(|\vec{x}|/t) e^{\lambda t} e^{- \frac{|\vec{x}|^2}{4 D t}}  \ . 
\end{eqnarray}
For $|\vec{x}|/t< v_*$, there is a suppression in the prefactor compared with~\eqref{stringotoc} 
%
%
\begin{equation}
\bigg|\frac{f_2(|\vec{x}|/t)}{f_1(|\vec{x}|/t)}\bigg| =  2 \sin\bigg(\frac{\pi d(d-1) \apr (1 - v^2/v_*^2)}{8 R^2} \bigg), \;\;\;\;\;\;\;\;\;\;\;\;\; v = |\vec{x}|/t
\label{commratio}
\end{equation}%
which can be through of as a higher dimensional counterpart of the $\cos(\lambda \beta_0/4)$ factor seen $0+1$ dimensional systems\footnote{Note $\cos(\lambda \beta_0/4) = \sin(\pi d(d-1) \apr/8R^2)$ and hence \eqref{commratio} reduces to $2 \cos(\lambda \beta_0/4)$ fior $v=0$.}. 

\section{A simple argument for the crossover velocity} \label{sec:velo}

As we have just discussed, in addition to the above example of holographic systems at finite $\apr$, it has been observed in various other systems that there can exist a crossover velocity $v_*$ beyond which maximally chaotic behavior is found even if theory is non-maximally chaotic for small $v$~\cite{Gu, Gu:2018jsv}. A more elaborate discussion of this crossover, and further examples of such systems, can be found in~\cite{Mezei:2019dfv}. In this section we consider a simplest scenario for the existence of such a crossover velocity\footnote{We thank discussion with Gabor Sarosi and Mark Mezei for the content of this section.}.

Consider higher spin exchange contribution to $H_4$ defined in~\eqref{otocholography}. At distances $|\vec{x}| \gg \b_0$ we expect 
the OTOC should decay exponentially with $|\vec{x}|$. In other words, the time ordering should not affect the validity of cluster decomposition principle at large distances. Denoting the contribution of a spin-$s$ operator 
to $H_4$ as $A_s$, we then expect 
\be \label{hdg}
A_s \propto e^{{2 \pi \ov \b_0} (s-1) t} e^{- M (s, \b_0) |\vec{x}|} 
\ee
where $M(s, \b_0)$ is a function of spin $s$ and the inverse temperature $\b_0$. $M(s, \b_0)$ will also depend on any coupling constants in theory (e.g. $\alpha'$ in string theory) and may also depend on other quantum numbers of the exchange operator. 
 For example, for a two-dimensional CFT in the large central charge limit, one finds that~\cite{Roberts:2014ifa} 
\be \label{hen1}
M (s, \b_0) = {2 \pi \ov \b_0} (\De - 1)
\ee
where $\De$ is the conformal dimension of the operator. 

Recall that the contribution of the stress tensor can be written as   
\be \label{hdg1}
A_{T}  \propto e^{{2 \pi \ov \b_0} (t - { |\vec x| \ov \tilde v_B})}  
\ee
where $\tilde v_B$ is not necessarily the butterfly velocity of the full system. 
Comparing~\eqref{hdg} with~\eqref{hdg1} we see that $A_s$ dominates over that from the stress tensor for a given $|\vec{x}|= v t$ if 
\be \label{hrn}
{2 \pi \ov \b_0} (s-1)  - M (s, \b_0) v  \geq {2 \pi \ov \b_0} \le(1 - { v \ov \tilde v_B} \ri) \ .
\ee
If~\eqref{hrn} is satisfied for one $s >2$ it must be 
satisfied for an infinite number of them, otherwise the chaos bound will be violated. Notice that the above equation is always satisfied for all $s> 2$ for $v=0$, which is consistent with the fact that upon including finite-coupling corrections SYK models and holographic systems will be non-maximally chaotic at $v=0$ \cite{Gu, Shenker2}.  

A simplest scenario for the existence a crossover velocity $v_*$ is the existence of a critical velocity $v_c$ such that 
for $v > v_c$, equation~\eqref{hrn} is not satisfied for any $s > 2$, while for $v < v_c$ it is satisfied for an infinite number of them. 
Then $v_c$ should provide an upper bound for $v_*$. 
Among all spin-$s$ operators, let us denote $M_0 (s, \b_0)$ as the smallest value for $M$. $M_0(s, \b_0)$ can be considered as defining a finite temperature version of the ``leading Regge trajectory.'' Then existence of such a $v_c$ then implies that 
\be 
{{2 \pi \ov \b_0} (s-2) \ov M_0 (s, \b_0) - {2 \pi \ov \b_0} {1 \ov \tilde v_B} } \leq v_c , \quad \forall s > 2 \ . 
\ee
In particular, the left hand side must have a finite limit as $s \to \infty$
\be 
 \lim_{s \to \infty}  {{2 \pi \ov \b_0} (s-2) \ov M_0 (s, \b_0) - {2 \pi \ov \b_0} {1 \ov \tilde v_B} } \equiv  v_\infty \leq v_c
\ee
$v_\infty$ can of course be larger or smaller than $v_*$. The above equation implies that $M_0(s, \b_0)$ should increase with $s$ at least as fast as linear dependence. In the case that $v_\infty$ is nonzero, $M_0$ must also be linear in $s$
\be 
\lim_{s \to \infty} M_0 (s, \b_0) = \ka (\b_0) s, \qquad  v_\infty = {2 \pi \ov \b_0} {1 \ov \ka (\b_0)} \ .
\ee 
We stress that the above discussion is only a simplest scenario. It can happen that $v_*$ exists without existence of $v_c$ or $v_\infty$.

\section{Discussion} \label{sec:conc}

In this paper we have discussed various features of a maximally chaotic system, motivated by the chaos effective field theory 
introduced in~\cite{Blake:2017ris}. Clearly an important future question is whether it is possible to write down an effective field theory for non-maximally chaotic systems. Away from maximal chaos, an infinite number of operators contribute to OTOCs. Thus the key is whether there exists a finite number of effective fields which can capture collective effects of the infinite number of operators. 
Right now there appears no obvious general guiding principle which enables us to identify these effective degrees of freedom or what should govern their dynamics. 
Important clues may come from Rindler OTOCs in a conformal field theory where 
non-maximal chaotic behavior arises in the Regge limit from resummation of contributions from infinite number of higher spin operators (see e.g.~\cite{Mezei:2019dfv} and references therein).

It is also of interest to better understand the physical interpretation of the vanishing of the leading order contribution to the commutator square for maximally chaotic systems.  It seems likely this vanishing indicates a sharp distinction between the nature of operator scrambling in maximally and non-maximally chaotic systems. One natural interpretation is that this vanishing suggests an extra ``unitarity'' constraint on the scrambling processes of maximally chaotic systems. Heuristically, if we view an OTOC as a scattering amplitude, the vanishing of the commutator square can be interpreted as the absence of ``particle'' production for the scattering process, i.e. the scattering is elastic. This can be made precise in various contexts. 
In the gravity description given in Appendix~\ref{app:scattering}, the phase shift for the corresponding string scattering is real in the maximally chaotic regime. Similarly, for Rindler OTOC, the vanishing of the commutator square means the vanishing of the imaginary part of the CFT scattering amplitude corresponding to the correlator~\cite{Caron-Huot:2017vep}. These precise statements resonate with the discussion of scrambling in generic 0+1 dimensional systems in~\cite{Kitaev,kitaIAS}, which used a restricted Hilbert space to define a scattering matrix that becomes unitary in the case of maximal chaos\footnote{In ~\cite{Kitaev,kitaIAS} this type of scrambling for maximal chaos has been referred to as being ``coherent''.}. Further, the existence of an enhanced ``unitarity'' for maximal chaos is related to the discussion of~\cite{Gu:2018jsv}, where the coefficient of exponential growth in commutator OTOCs is inversely proportional to the branching time.  As such this suggests that branching processes are absent in systems for which commutator squares vanish.

\section*{Acknowledgements}
We would like to thank Gabor Sarosi and Mark Mezei for helpful discussions. This work is supported by the Office of High Energy Physics of U.S. Department of Energy under grant Contract Number  DE-SC0012567.

\appendix

\section{Formula for commutator square in holographic systems}
\label{app:scattering} 

In this Appendix we present an expression for the commutator square in holographic systems which is a simple generalization of 
the formula for OTOC in terms of 
the eikonal approximation to the bulk scattering process presented in~\cite{Shenker2}.  We will only briefly explain various notations below, and refer the reader to \cite{Shenker2} for further details. We will consider a slightly more general set of correlation functions than discussed in Section~\ref{sec:holography}, 
\begin{eqnarray}
&H_4&(\{{t_1, \vec{x}_1}\},\{{t_2+ i \theta, \vec{x}_2}\},\{{t_3, \vec{x}_3}\},\{{t_4 + i \theta, \vec{x}_4}\}) = \langle W_{\vec{x}_4} (t_4 + i \theta) V_{\vec{x}_2}(t_2 + i \theta) W_{\vec{x}_3} (t_3) V_{\vec{x}_1}(t_1) \rangle \nonumber \\
&\tilde{H}_4&(\{{t_1, \vec{x}_1}\},\{{t_2 + i \theta, \vec{x}_2}\},\{{t_3, \vec{x}_3}\},\{{t_4 + i \theta, \vec{x}_4}\}) = \langle W_{\vec{x}_4} (t_4 + i \theta) V_{\vec{x}_2}(t_2 + i \theta) W_{\vec{x}_3} (t_3) V_{\vec{x}_1}(t_1) \rangle \nonumber \\
&C_4^{\theta}& (\{{t_1, \vec{x}_1}\},\{{t_2, \vec{x}_2}\},\{{t_3, \vec{x}_3}\},\{{t_4, \vec{x}_4}\}) = \langle [W_{\vec{x}_4} (t_4 + i \theta), V_{\vec{x}_2}(t_2 + i \theta)][ W_{\vec{x}_3} (t_3), V_{\vec{x}_1}(t_1) ]\rangle \nonumber 
\end{eqnarray}
%
%
Recall that  $H_4$ defined above can be expressed as~\cite{Shenker2}
\begin{equation}
\label{amplitudeapp}
H_4(\{t_i,\vec{x}_i\}) = \frac{a_0^4}{(4 \pi)^2} \int e^{i \delta(s, b)} \bigg[p_{1}^{u} \psi_{4}^{*}(p_1^{u}, \vec{x}) \psi_{3}(p_{1}^{u},\vec{x}) \bigg] \bigg [p_{2}^{v} \psi_2^{*} (p_2^{v}, \vec{x}') \psi_1(p_2^{v}, \vec{x}') \bigg]
\end{equation}
where the integral runs over $\vec{x}, \vec{x}', p^{u}_1, p^{v}_2$ and the parameter $a_0$ is simply a constant. Further, $\delta(s,b)$ is the phase shift of a bulk two-to-two scattering process and is expressed as a function of the centre of mass energy $s = a_0 p^{u}_1 p^{v}_2 $ and transverse separation $b = |\vec{x} - \vec{x}'|$ of the scatterred quanta. The various $\psi_i$ in \eqref{amplitudeapp} are Fourier transforms of bulk-to-boundary propagators\footnote{Note that the labelling of fields and operators in \eqref{amplitudeapp} is slightly different to the corresponding formula in \cite{Shenker2} as we are considering a slightly different arrangement of operators.}
\begin{eqnarray}
 \psi_1(p^v,\vec{x}) &=& \int du e^{i a_0 p^{v} u/2} \langle \phi_{V}(u,v,\vec{x}) V_{\vec{x}_1}(t_1) \rangle \bigg|_{v=0}\nonumber \\ 
\psi_2(p^v,\vec{x}) &=& \int du e^{i a_0 p^{v} u/2} \langle \phi_{V}(u,v,\vec{x}) V_{\vec{x}_2}(t_2 - i \theta)\rangle  \bigg|_{v=0}\nonumber \\
\psi_3(p^u,\vec{x}) &=& \int dv e^{i a_0 p^{u} v/2} \langle \phi_{W}(u,v,\vec{x}) W_{\vec{x}_3}(t_3) \rangle  \bigg|_{u=0}\nonumber \\
\psi_4(p^u,\vec{x}) &=& \int dv e^{i a_0 p^{u} v/2} \langle \phi_{W}(u,v,\vec{x}) W_{\vec{x}_4}(t_4 - i \theta) \rangle  \bigg|_{u=0}
\label{eq:bulkboundary}
\end{eqnarray}
with $\phi_V,\phi_W$ the dual bulk fields dual to Hermitian boundary operators $V, W$.  Using the identity 
\begin{equation}
\tilde{H}_4(\{t_1,\vec{x}_1\},\{t_2 + i \theta ,\vec{x}_2\},\{t_3,\vec{x}_3\},\{t_4 + i \theta ,\vec{x}_4\})^* = H_4(\{t_2 - i \theta ,\vec{x}_2\},\{t_1,\vec{x}_1\},\{t_4 - i \theta ,\vec{x}_4\},\{t_3,\vec{x}_3\}) \nonumber 
\end{equation}
to compute $\tilde{H}_4$ we find that the commutator $C_4^{\theta} = H_4 + \tilde{H}_4 + \dots$ can be written as \footnote{Here we are only interested in the exponential growth of $C_4^{\theta}$, so are ignoring the time-ordered pieces}
\begin{equation}
\label{amplitudeapp2}
 C_4^\theta(\{t_i,\vec{x}_i\})= \frac{a_0^4}{(4 \pi)^2} \int (e^{i \delta(s,b) }+ e^{-i \delta^*(s, b)}) \bigg[p_{1}^{u} \psi_{4}^{*}(p_1^{u}, \vec{x}) \psi_{3}(p_{1}^{u},\vec{x}) \bigg] \bigg [p_{2}^{v} \psi_2^{*} (p_2^{v}, \vec{x}') \psi_1(p_2^{v}, \vec{x}') \bigg]
\end{equation} 
with the bulk to boundary propagators again given by the formulae in \eqref{eq:bulkboundary}. In both string theory and classical gravity  $\delta \sim G_N \sim 1/{\cal N}$, and hence an expansion in $1/{\cal N}$ amounts to expanding the exponentials in \eqref{amplitudeapp} and \eqref{amplitudeapp2}. To leading order $1/{\cal N}$ we therefore find 
\begin{eqnarray}
\label{amplitude3}
H_4(\{t_i,x_i\}) = \frac{i a_0^4}{(2 \pi)^4} \int \delta(s,b) &\bigg[&p_{1}^{u}  \psi_{1}^*(p_{1}^{u},x) \psi_{3}(p_1^{u}, x)\bigg] \bigg [p_{2}^{v} \psi_2^*(p_2^{v}, x')  \psi_4 (p_2^{v}, x') \bigg] \nonumber \\
 C_4^\theta(\{t_i,\vec{x}_i\})  = -\frac{2 a_0^4}{(2 \pi)^4} \int \mathrm{Im} \delta(s,b) &\bigg[&p_{1}^{u}  \psi_{1}^*(p_{1}^{u},\vec{x}) \psi_{3}(p_1^{u}, \vec{x})\bigg] \bigg [p_{2}^{v} \psi_2^*(p_2^{v}, \vec{x}')  \psi_4 (p_2^{v}, \vec{x}') \bigg] \ .
\end{eqnarray}
For theories dual to classical gravity $\delta(s,b)$ is always purely real and can be computed by evaluating the on-shell gravitational action of a pair of gravitational shock waves \cite{Roberts,Shenker2}. It is immediately clear from \eqref{amplitude3} that all $C_\theta$ vanish identically to leading order in $1/{\cal N}$ in such theories within the eikonal approximation. 

In string theory the phase shift $\delta(s,b)$ obtains an imaginary part. For the black-hole background discussed in \cite{Shenker2} the phase shift $\delta(s,b)$ was found to take the form
\begin{equation}
\label{stringphaseshift}
\delta(s,b) \sim G_N s \int \frac{d^{d-1} k}{(2 \pi)^{d-1}} \frac{e^{i \vec{k} \cdot \vec{y}}}{\vec{k}^2 + \mu^2} ( e^{- i \pi/2} \alpha' s/4)^{-\alpha'(\vec{k}^2 + \mu^2)/2r_0^2}, \quad s = c_1e^{2 \pi t/\beta_0} 
\end{equation}
where $b  = |\vec{y}|$, $d-1$ is the number of boundary space dimensions, $\mu$ is a constant parameter related to the butterfly velocity, $r_0$ is the horizon radius and $\alpha'= l_s^2$ (with $l_s$ the string length). We thus find
\begin{equation}
\label{stringphaseshift2}
\mathrm{Im} (\delta(s,b) )\sim G_N s \int \frac{d^{d-1} k}{(2 \pi)^{d-1}} \frac{e^{i \vec{k} \cdot \vec{y}}}{\vec{k}^2 + \mu^2}  \mathrm{sin}({ \pi \alpha'(\vec{k}^2+ \mu^2)/4 r_0^2})( \alpha' s/4)^{-\alpha'(\vec{k}^2 + \mu^2)/2r_0^2} \ .
\end{equation}
To see the behavior of~\eqref{stringphaseshift2} let us first recall the evaluation of~\eqref{stringphaseshift}~\cite{Shenker2}. 
With $\vec{y} \equiv \vec{v} t$ and the explicit form of $s$ plugged in, equation~\eqref{stringphaseshift} can be written as 
\begin{equation}
\label{stringphaseshift3}
\delta(t, v) \sim G_N  c_1 \int \frac{d^{d-1} k}{(2 \pi)^{d-1}} \frac{e^{i \vec{k} \cdot \vec{v} t + \frac{2 \pi t}{\beta_0}(1 - {B(\vec{k})})}}{\vec{k}^2 + \mu^2} ( c_1 e^{- i \pi/2} \alpha'/4)^{-B(\vec{k})}, \;\;\;\;\;\; B(\vec{k}) = \alpha'(\vec{k}^2 + \mu^2)/2r_0^2, 
\end{equation}
where $v = |\vec{v}|$. At large $t$ the integral in \eqref{stringphaseshift3} has a saddle-point at $\vec{k}_*$ where
\begin{equation} 
i \vec{v} = \frac{2 \pi}{\beta_0} \frac{\partial B}{\partial \vec{k}}\bigg|_{\vec{k}_*} \;\;\;\; \implies \vec{k}_* 
= {i \vec{v} \mu \ov v_*} , \quad  v_* \equiv \frac{2 \pi \alpha' \mu}{\beta_0 r_0^2} \ .
\end{equation}
The integrand in \eqref{stringphaseshift3} also has a pole at 
\begin{equation}
\vec{k}^2 + \mu^2 = 0, \;\;\;\;\;\;\; \mu = \frac{2 \pi}{\beta_0 v_B} 
\end{equation}
with $v_B$ the butterfly velocity.  For $v < v_*$ the integral can be computed using the saddle-point approximation. The resulting expression for the phase-shift $\delta(t,v)$ evaluates to   
\begin{equation} \label{hen}
\delta(t,v) \sim \frac{1}{(4 \pi D t)^{(d-1)/2}} \tilde{f}_1(v) e^{\lambda(v) t} 
\end{equation}
where $\lambda(v)$ is the velocity dependent Lyapunov exponent \cite{Khemani:2018sdn, Gu:2018jsv}
\begin{equation}
\lambda(v) =  \frac{2 \pi}{\beta_0}\bigg(1 - \frac{d(d-1)\alpha'}{4 R^2} \bigg) - \frac{v^2}{4 D},   \;\;\;\;\;\; D = \frac{d^2 \alpha' \beta_0}{16 \pi R^2} 
\label{diffusionVDLE}
\end{equation}
and
\begin{equation}
\tilde{f}_1(v) = \frac{G_N c_1 \beta_0^2 d}{8 \pi^2 (d-1) }\frac{1}{(1 - v^2/v_*^2)} (c_1 e^{- i \pi/2} \alpha'/4)^{ -\frac{d(d-1)\alpha'}{4 R^2} (1 - v^2/v_*^2)} 
\end{equation}
where in obtaining these expressions we have eliminated $\mu$ and $r_0$ using that for the black hole background studied in \cite{Shenker2} we have $v_B  = \sqrt{d/2(d - 1))}$ and $r_0 =  4 \pi R \beta_0^{-1} d^{-1}$ with $R$ the AdS radius. Recalling that the bulk to boundary wavefunctions in \eqref{amplitudeapp} are spatially peaked around the positions of the external operators we have from \eqref{amplitude3} that the functional dependence of \eqref{otocholography} is given at leading order in $1/{\cal N}$ by $i \delta(t, v)$ with $v = |\vec{x}|/t$. This gives rise to the functional form of \eqref{stringotoc} discussed in the main text. 

Note that the expression~\eqref{hen} diverges at $v= v_*$, which can be attributed to the existence of 
a graviton pole at $\vec{k}^2 + \mu^2 = 0$ in the integrand of~\eqref{stringphaseshift3}. For $v > v_*$ 
the contribution from the graviton pole dominates and leads to the behavior of maximal chaos. 

In contrast, there is no graviton pole in~\eqref{stringphaseshift2} due to the factor of $\mathrm{sin}({ \pi \alpha'(\vec{k}^2 + \mu^2)/4 r_0^2})$ in the numerator. So in this case, the saddle-point approximation is valid for all $\vec{x}$ and $t$. We find 
\begin{equation}
\Im(\delta(t ,v)) = \frac{1}{(4 \pi D t)^{(d-1)/2}} \tilde{f}_2(v) e^{\lambda(v) t} 
\end{equation}
where now 
\begin{equation}
\tilde{f}_2(v) =  \frac{G_N c_1 \beta_0^2 d}{8 \pi^2 (d-1)  }\frac{1}{(1 - v^2/v_*^2)} (c_1 \alpha'/4)^{ -\frac{d(d-1)\alpha'}{4 R^2} (1 - v^2/v_*^2)}  \sin \bigg( \frac{ \pi d(d-1)\alpha'}{8 R^2}(1 - v^2/v_*^2)\bigg)   
\end{equation}
Again from \eqref{amplitude3} the functional dependence of the commutator square in \eqref{commutatorcholography} is given by $2 \Im(\delta(t , v))$ with $v = |\vec{x}|/t$, thus giving rise to \eqref{oen} and \eqref{commratio}.

\end{document}